\documentclass[aps,prl,twocolumn,superscriptaddress,showpacs,preprintnumbers,amsmath,amssymb]{revtex4-2}
\usepackage{graphicx}% Include figure files
\usepackage{dcolumn}% Align table columns on decimal point
\usepackage{bm}% bold math
\usepackage{txfonts}
\usepackage{float}
\usepackage{latexsym}

\usepackage{color}
\usepackage[normalem]{ulem}
\begin{document}

%\preprint{APS/123-QED}

\title{Effects of vortex and antivortex excitations in underdoped Bi$_2$Sr$_2$Ca$_2$Cu$_3$O$_{10+\delta}$ bulk single crystals} % Force line breaks with \\
%\title{EFFECTS OF VORTEX AND ANTI-VORTEX EXCITATIONS IN UNDERDOPED BI-2223 BULK SINGLE CRYSTALS } 

\author{Takao Watanabe}
\email{Present address: Physics Department, College of Engineering, Nihon University, Fukushima 963-8642, Japan. E-mail address: 
watanabe.takao@nihon-u.ac.jp}
\affiliation{Graduate School of Science and Technology, Hirosaki University, Hirosaki, Aomori, 036-8561 Japan}

\author{Kenta Kosugi}
\affiliation{Graduate School of Science and Technology, Hirosaki University, Hirosaki, Aomori, 036-8561 Japan}
	\author{Nae Sasaki}
\affiliation{Graduate School of Science and Technology, Hirosaki University, Hirosaki, Aomori, 036-8561 Japan}
	\author{Shunpei Yamaguchi}
\affiliation{Graduate School of Science and Technology, Hirosaki University, Hirosaki, Aomori, 036-8561 Japan}
	%\affiliation{Graduate School of Science and Technology, Hirosaki University, 3 Bunkyo, Hirosaki, 036-8561 Japan}
	%\author{Mihaly M. Dobroka$^1$}
	%\affiliation{Graduate School of Science and Technology, Hirosaki University, 3 Bunkyo, Hirosaki, 036-8561 Japan}
%	\author{Shintaro Adachi}
% \affiliation{Nagamori Institute of Actuators, Kyoto University of Advanced Science (KUAS), Kyoto 615-8577, Japan}
        \author{Takenori Fujii}
\affiliation{Cryogenic Research Center, University of Tokyo, Bunkyo, Tokyo 113-0032, Japan}
\author{Ken Hayama}
\affiliation{Department of Electronic Science and Engineering, Kyoto University, Kyoto 615-8510, Japan}
	\author{Itsuhiro Kakeya}
\affiliation{Department of Electronic Science and Engineering, Kyoto University, Kyoto 615-8510, Japan}

        \author{Toshimitsu Ito}
\affiliation{Research Institute for Advanced Electronics and Photonics, National Institute of Advanced Industrial Science and Technology (AIST), Higashi 1-1-1, Tsukuba, Ibaraki 305-8565, Japan}
%        \author{Shojiro Kimura}
% \affiliation{Institute for Materials Research, Tohoku University, 2-1-1 Katahira, Aoba-ku, Sendai, 980-8577 Japan}
%	\author{Ken Hayama}
%\affiliation{Department of Electronic Science and Engineering, Kyoto University, Kyoto 615-8510, Japan}

	%\affiliation{Institute for Solid State Physics, University of Tokyo, 5-1-5 Kashiwanoha, Kashiwa, Chiba 277-8581, Japan}
	%\author{Koichi Kindo$^3$}
	%\affiliation{Institute for Solid State Physics, University of Tokyo, 5-1-5 Kashiwanoha, Kashiwa, Chiba 277-8581, Japan }
	%\author{Hajime Ishikawa}
%\affiliation{Institute for Solid State Physics, University of Tokyo, Kashiwa, Chiba 277-8581, Japan}
        %\author{Koichi Kindo}
%\affiliation{Institute for Solid State Physics, University of Tokyo, Kashiwa, Chiba 277-8581, Japan}
       % \author{Daniel S. Dessau}
% \affiliation{Department of Physics, University of Colorado at Boulder, Boulder, CO 80309, USA}

	%\affiliation{Institute for Materials Research, Tohoku University, 2-1-1 Katahira, Aoba-ku, Sendai, 980-8577 Japan}
%	\author{Takao Watanabe}
%\email{twatana@hirosaki-u.ac.jp}
%\affiliation{Graduate School of Science and Technology, Hirosaki University, Hirosaki, Aomori, 036-8561 Japan}
	%\affiliation{Graduate School of Science and Technology, Hirosaki University, 3 Bunkyo, Hirosaki, 036-8561 Japan}
%$^\star$\thanks{email: twatana@hirosaki-u.ac.jp}

	%\affiliation{Institute for Solid State Physics, University of Tokyo, 5-1-5 Kashiwanoha, Kashiwa, Chiba 277-8581, Japan$^3$}

%\author{Aaa Bee}

%\affiliation{Hirosaki University, Japan}


\date{\today}% It is always \today, today,

\begin{abstract}
The observance of vortex and anti-vortex effects in bulk crystals can prove the existence of phase-disordered superconductivity in the bulk. To gain insights into the mechanisms that govern superconducting transition in copper oxide high-transition temperature ($T_c$) superconductors, this study investigated the transport properties of underdoped Bi$_2$Sr$_2$Ca$_2$Cu$_3$O$_{10+\delta}$ (Bi-2223) bulk single crystals. The $I$–$V$ characteristics results and the typical tailing behavior owing to the temperature dependence of in-plane resistivity ($\rho_{ab}$) were consistent with the  Kosterlitz-Thouless (KT) transition characteristics. Thus, with increasing temperature, copper oxide high-$T_c$ superconductors transitioned to their normal state owing to destruction of their phase correlations, although a finite Cooper pair density was prevalent at $T_c$. Further, magnetization measurements were performed to determine the temperature dependence of the irreversible magnetic field $B_{irr}$. Consequently, the mechanism governing the KT transition-like superconducting transition in this bulk system was elucidated. These results support the extreme strong-coupling models for high-$T_c$ superconductivity in cuprates.
%Using the high-quality multilayered Bi-2223 single crystals, the intriguing superconducting transition that strongly support the extreme strong-coupling scenario for high-$T_c$ superconductivity have been demonstrated.

%\begin{description}
%\end{description}
\end{abstract}

%\pacs{71.27.+a, 79.60.-i}% PACS, the Physics and Astronomy
\maketitle

\section{I. INTRODUCTION}%%%INTRODUCTION%%%
In conventional Bardeen, Cooper, and Schrieffer (BCS) superconductors, the amplitude of the order parameter disappears as the temperature increases till the superconducting transition temperature ($T_c$) is reached. However, in copper-oxide high-$T_c$ superconductors, the order parameter phase is destroyed despite its finite amplitude, which resuls in a transition to a normal state \cite{Uemura89, Kivelson95, Franz98}. Thus, a "phase-disordered superconductivity" is observed at temperatures higher than $T_c$. However, the plausibility of this theory must be investigated to elucidate the mechanism of the high-$T_c$ superconductivity in cuprates. Despite the several interesting experimental results reported \cite{Ong00, Ong05} in previous studies, there is no consensus among researchers. 

In two-dimensional (2D) superconductors, at high temperatures approaching $T_c$, the free vortices (circulating super-currents are counterclockwise) and anti-vortices (circulating super-currents are clockwise) are thermally excited in equal numbers. Cosequently, the long-range phase ordering of superconductivity is disrupted. However, below a certain temperature ($T_{KT}$), vortex-anti-vortex pairs are formed individually, which undergo a phase transition to a substantial superconducting state. This is referred to as the Kosterlitz-Thouless (KT) transition \cite{Kosterlitz73, Beasley79}. Therefore, the observance of vortex and anti-vortex effects in bulk crystals can prove the existence of phase-disordered superconductivity in the bulk \cite{Franz07}.

Several studies have focused on the KT transition in cuprate high-$T_c$ superconductors, on 1-unit-cell-thick YBa$_2$Cu$_3$O$_{7-\delta}$ (YBCO) ultra-thin films \cite{Matsuda92}, 2-unit-cell Ca-doped YBCO ultrathin films \cite{Hetel07}, and 2-unit-cell Bi$_2$Sr$_2$Ca$_2$Cu$_3$O$_{10+\delta}$ (Bi-2223) that were mechanically exfoliated from the bulk crystal \cite{Yu22}. All these studies used 2D films. However, it is considered that the KT transition does not occur in a bulk system with a finite interlayer coupling. This is because, it is a unique phase transition in a 2D system. Certain studies have focused on the possible KT transition; for example, La$_{1.875}$Ba$_{0.125}$CuO$_4$ bulk single crystals \cite{Tranquada07} and underdoped La$_{2-x}$Sr$_x$CuO$_4$ (LSCO) thick films (equivalent to the bulk) \cite{Kitano06}. However, Matsuda {\it et al.} found \cite{Matsuda93} that even in Bi$_2$Sr$_2$CaCu$_2$O$_{8+\delta}$ (Bi-2212), which exhibits the strongest 2D nature among copper oxides, the occurrence of the KT transition is challenging owing to the system obeying a simple three-dimensional (3D)-XY model \cite{Hikami80}.

Multilayered high-$T_c$ cuprates are suitable for exploring the KT transition in bulk because the inner CuO$_2$ planes are ideally flat and underdoped compared to the outer CuO$_2$ planes\cite{Mukuda12, Kunisada20}. Thus, they are expected to decouple the interplane Josephson coupling \cite{Iye10, Nomura19}. Using underdoped samples of trilayered Bi-2223, for which good quality single crystals are readily available \cite{Fujii01, Adachi15}, this study examined the possibility of the KT transition. The $I$-$V$ characteristics and tailing behavior of $\rho_{ab} (T)$ indicated the occurrence of a KT transition-like superconducting transition. Furthermore, based on the measurements of the irreversible magnetic field $B_{irr} (T)$, we discussed the mechanism that enables KT transition-like vortex and anti-vortex excitations in bulk materials.

\begin{figure*}[t]
		\begin{center}
			\includegraphics[width=180mm]{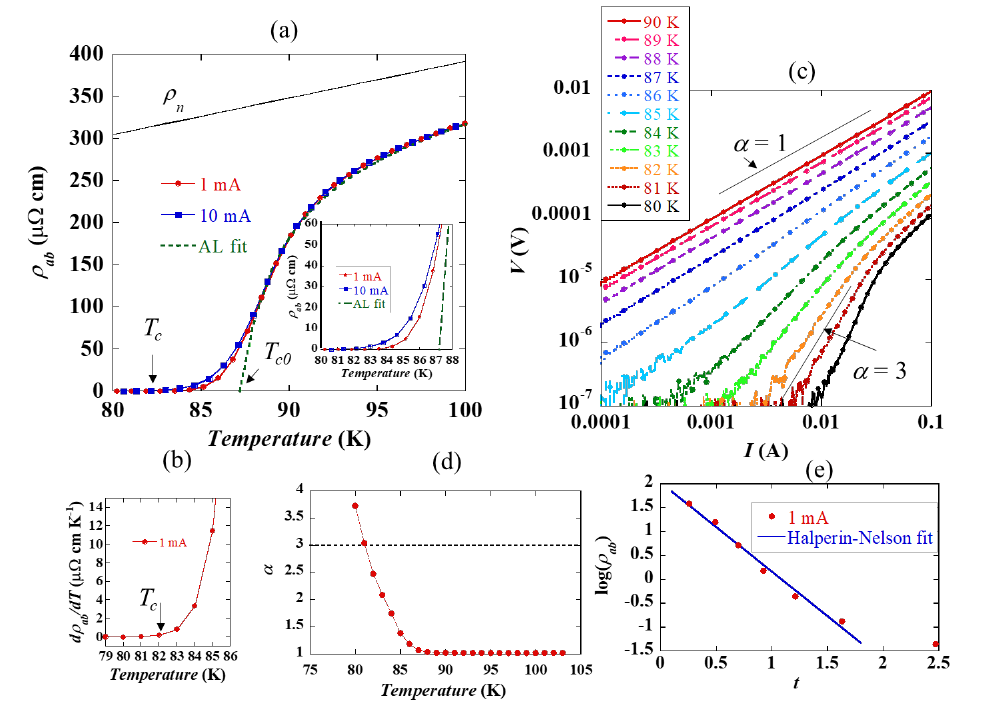}
			\caption{\label{fig1}(Color online) Kosterlitz-Thouless (KT)-like superconducting transition in an underdoped (UD1) Bi-2223 bulk single crystal. (a) Superconducting transition curve for a UD1 Bi-2223 single crystal. The green dashed line shows the fitting results  using the 2D Aslamasov—Larkin (AL) formula for superconducting fluctuations. The mean field superconducting transition temperature ($T_{c0}$) is estimated to be 87 K. $\rho_{n}$ is the in-plane resistivity ($\rho_{ab}$) in the absence of superconducting fluctuations used for the fitting, and $\rho_{n}$ is assumed to be $\rho_{n}$ = $aT+b$ ($a$ = 4.4 $\mu\Omega$cm/K, $b$ = -55 $\mu\Omega$cm). The inset shows expanded plots for non-ohmic behavior observed below 87 K. (b) The temperature dependence of $d\rho_{ab} (T)/dT$. From this, $T_c$ is obtained as 82 K. (c) $I$-$V$ characteristics at various temperatures plotted on both axes on the logarithmic scales. (d) Power exponents $\alpha$ found in the $I$-$V$ characteristics ($V \propto I^{\alpha}$) as a function of temperature. $\alpha$ reaches a value of 3 at 81 K [the KT transition temperature ($T_{KT}$) = 81 K]. (e) Temperature dependence of $\rho_{ab}$. The straight line indicates the fitted result by the Halperin--Nelson equation, $\rho_{ab} (T)$ = $\rho_{ab}^0$$\exp(-2ct)$, where $t = [({T_{c0} - T})/(T - T_{KT})]^{1/2}$  ($\rho_{ab}^0$ = 106 $\mu\Omega$cm and $c$ = 2.15).}  
		\end{center}
	\end{figure*}

\section{II. EXPERIMENT}
High-quality Bi-2223 single crystals were grown employing the traveling solvent floating zone (TSFZ) method \cite{Adachi15}. The raw rod composition was Bi-rich (Bi:Sr:Ca:Cu = 2.25:2:2:3), the growth atmosphere was an oxygen-argon mixture with 10 \% oxygen, and the growth rate was set to 0.05 mm/h. The crystals obtained were annealed at 600 $^\circ$C for 4 h at an oxygen partial pressure ($P_{O2}$) of 5 or 2 Pa to facilitate underdoping. Hereinafter, the crystals annealed under these conditions are referred to as UD1 Bi-2223 and UD2 Bi-2223. 

To obtain the $I$–$V$ characteristics, pulse measurements were performed to minimize heat generation under a high current bias. A pulse period and width of 50 and 1 ms (thus, a duty ratio of 2 \%), respectively, were used (Appendix A). The in-plane resistivity $\rho_{ab} (T)$ was measured using the DC 4-terminal method. Further, the magnetic susceptibility was measured using a superconducting quantum interference device magnetometer (Quantum Design Magnetic Property Measurement System). Irreversible magnetic fields ($B_{irr}$) were obtained through measurements of the temperature dependence of magnetic susceptibility under various magnetic fields ($B \parallel c$) up to 7 T for both zero-field cooling (ZFC) and cooling in a magnetic field (FC). All the temperature sweep rates were set as 1 K/min.

\section{III. RESULTS}
Figure 1(a) shows the $\rho_{ab} (T)$ of the underdoped ($T_c$ = 82 K) Bi-2223 single crystal (UD1 Bi-2223) near $T_c$ (the results of UD2 Bi-2223 are presented in the Appendix B).
Below 100 K, a largely rounded temperature-dependence characteristic of Bi-based copper oxides \cite{Adachi151} was observed, whereas tailed behavior was observed  at temperatures below 87 K. As $\rho_{ab} (T)$ gradually approached zero, $T_c$ was defined as the temperature at which $d\rho_{ab} (T)/dT$ became negligibly small ($d\rho_{ab} (T)/dT$ $\approx$ 0.2 $\mu\Omega$cm/K) (Fig. \ref{fig1} (b)). The largely rounded temperature dependence was attributed to superconducting fluctuations. This dependence was further analyzed using the formula $\rho_{ab} (T)$ = 1/($\rho_{n} (T)^{-1}$ + $\sigma_{2D-AL} (T)$), where $\rho_{n} (T)$ is the in-plane resistivity in the absence of superconducting fluctuation effects. It is assumed to be $\rho_{n} (T)$ = $aT$ + $b$ ($a$ and $b$ are constants). $\sigma_{2D-AL} (T)$ is the excess conductivity associated with the 2D Aslamazov—Larkin (AL) superconducting fluctuation \cite{Aslamasov68}. [Here, $\sigma_{2D-AL} (T)$ = $e^2\epsilon^{-1}$/16$\hbar$$s$, where $\epsilon$ is the reduced temperature, $\epsilon$ = $ln(T/T_{c0})$, $T_{c0}$ is the mean field, $T_c$, and $s$ (= 18.5 \AA) is the distance between the conductive planes.] 

At temperatures above 88 K, the fitting reproduced the experimental results well. Thus, $T_{c0}$ was determined as 87 K with a dissociation of 5 K at $T_c$ (= 82 K). Furthermore, when the applied current was increased to 10 mA, the tailed behavior was enhanced in magnitude at approximately 85 K (inset in Fig. \ref{fig1} (a)). These results were not attributable to the inhomogeneity of the sample (see Appendix C). Thus, these results implied that the tailed behavior of the in-plane resistivity was owing to a deviation from the mean-field theory, that is, the superconducting fluctuations. The most likely cause was the flux flow resistance owing to the excitation of spontaneous vortices and anti-vortices associated with the KT transition. 

The excited free vortex and anti-vortex were all paired at the transition temperature of the KT transition, $T_{KT}$, such that the electrical resistance was zero at the limit of the zero bias current. However, under a finite bias current, the vortex and the antivortex were pulled apart in opposite directions by the Lorentz force. Consequently, the number $n_{free}$ of free vortices and anti-vortices became finite and resistance was generated. In this case, the relationship between $n_{free} \propto I^{\pi{K}}$ was derived \cite{Kadin83}. Here, $K$ is a constant proportional to the effective superfluid sheet density of the material. Therefore, the following $I$–$V$ characteristics can be obtained from resistance $( =V ⁄I) \propto n_{free}$,
$V \propto I^{\alpha}$ with $\alpha = 1+ \pi{K}$.
At $T = T_{KT}$, $K$ is known to jump from zero to $2⁄\pi$ upon cooling. Therefore, at $T = T_{KT}$, the power exponent $\alpha$ found in the $I$–$V$ characteristic jumps to three \cite{Kadin83}. This universal jump in the value of $\alpha$ is a phenomenon peculiar to KT transition.

Figure 1(c) shows a log-log plot of the $I$–$V$ characteristics measured at various temperatures 80-90 K. The data on the low-current side exhibited a linear slope, indicating $V \propto I^{\alpha}$. The power exponent $\alpha$ was obtained from the slope of the graph. The obtained $\alpha$ was plotted against temperature (Fig. \ref{fig1} (d)). Above 88 K, the voltage ($V$) was proportional to the current ($I$) (i.e., $\alpha$ = 1). However, below 87 K, $\alpha$ increased slightly from 1. This is probably because below $T_{c0}$, free-vortex and anti-vortex excitations occur, some of which pair up and are dissociated by the Lorentz force with increase in the applied current. When the temperature was further reduced, $\alpha$ rapidly increased. Subsequently, $\alpha$ reached three at 81 K, which was slightly lower than $T_c$ (= 82 K). Thus, we attributed this non-linear $I$–$V$ characteristics to the KT transition, and defined $T_{KT}$ of this sample as 81 K ($T_{KT}$ = 81 K). To the best of our understanding, no other explanation may be possible for this observation (Appendix D). However, immediately prior to the occurrence of the KT transition (at a temperature slightly higher than $T_{KT}$), the system may have been subjected to a normal 3D superconducting transition owing to interplanar interactions. In fact, a finite superconducting current flowed below $T_c$ (see Fig. \ref{figS6} in the Appendix E).

The KT theory predicts the existence of free vortices and antivortices at $T$ $>$ $T_{KT}$, which facilitates the associated vortex flow resistance. Owing to the vortex flow resistance being proportional to the density of free vortices and anti-vortices, Halperin and Nelson \cite{Halperin79} proposed the following equation, $\rho_{ab} (T)$ = $\rho_{ab}^0$$\exp(-2ct)$,
%\begin{equation}
%\[
%\rho_{ab} (T) = \rho_{ab}^{0}\exp\biggl[-2c \bigg(\frac{T_{c0} - T}{T - T_{KT}}\bigg)^{1/2}\biggr], 
%\]
%\end{equation}
where $c$ is a constant of the order of 1 and $t = [({T_{c0} - T})/(T - T_{KT})]^{1/2}$. Figure 1(e) shows the $\rho_{ab} (T)$ as a function of $t$ for an applied current of 1 mA. The data were almost on a straight line, indicating that the origin of the tailed $\rho_{ab} (T)$ below 87 K was attributable to the excitation of the free vortices and antivortices.

\begin{figure}[t]
		\begin{center}
			\includegraphics[width=90mm]{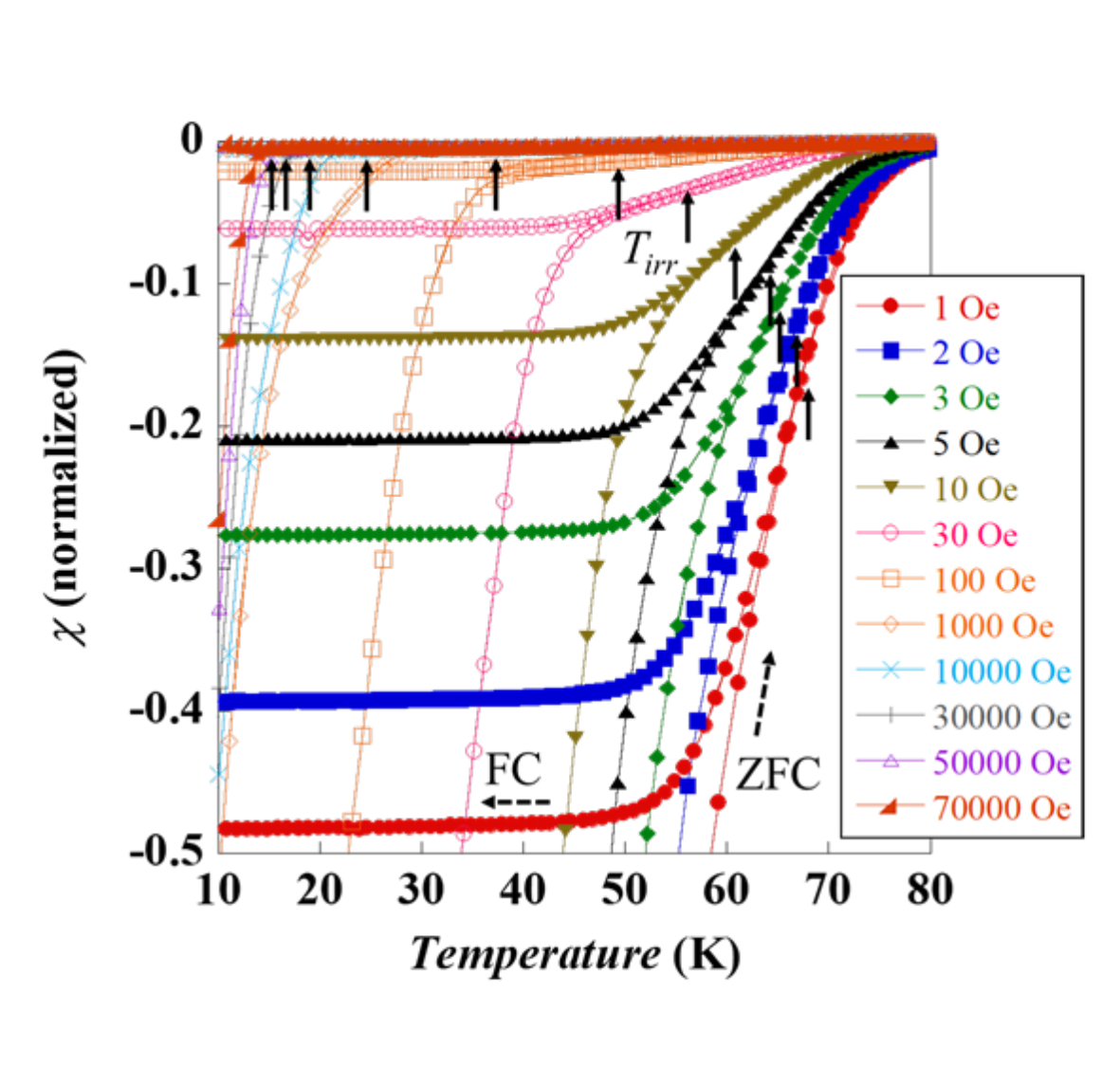}
			\caption{\label{fig2}(Color online)Temperature dependence of magnetic susceptibility ($\chi$) of underdoped (UD) Bi-2223 bulk single crystal under various magnetic fields ($B \parallel c$).  Data for the zero--field cooling (ZFC) and cooling in a magnetic field (FC) are indicated via dashed arrows (for 1 Oe data only). $\chi$ is normalized by the ZFC value at the lowest temperature. Solid arrows indicate the irreversible temperature ($T_{irr}$) under the respective magnetic field.}
		\end{center}
	\end{figure}

\begin{figure}[t]
		\begin{center}
			\includegraphics[width=90mm]{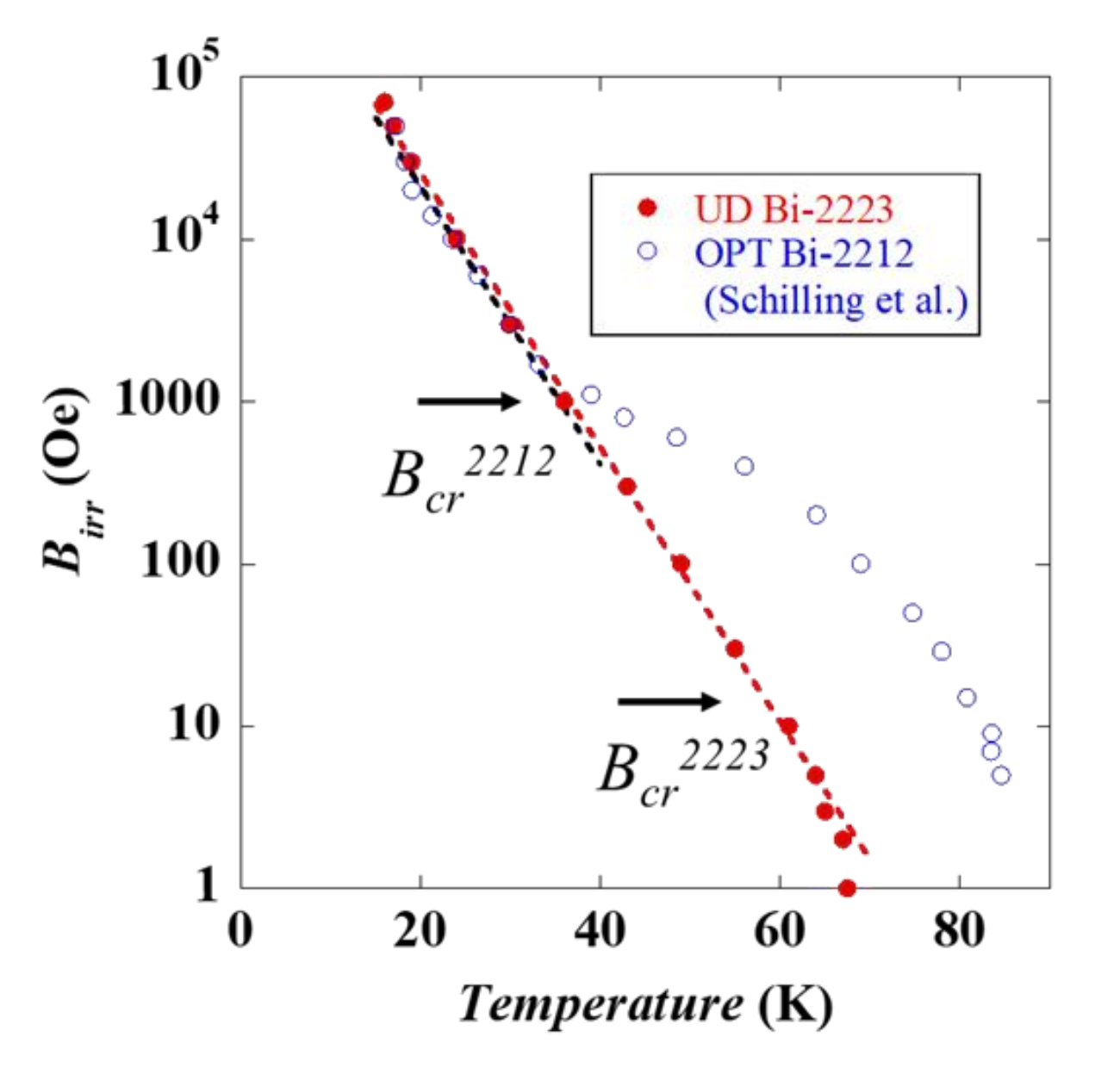}
			\caption{\label{fig3}(Color online)Temperature dependence of the irreversible magnetic field ($B_{irr}$) for underdoped (UD) Bi-2223 and optimally doped (OPT) Bi-2212 \cite{Schilling93}. The vertical axis is shown on a logarithmic scale. The dotted lines indicate the fitting results using the formula, $B_{irr} (T) = B_{0}e^{-T/T_{0}}$, in the low-temperature and high-field region. The crossover fields ($B_{cr}$) from two to three dimensions are indicated via the arrows for OPT Bi-2212 and UD Bi-2223.}
		\end{center}
	\end{figure}

To generate the vortex/anti-vortex states, the electronic system must be extremely 2D in nature. To confirm this, this study investigated the temperature dependence of the magnetic susceptibility ($\chi$) of underdoped single crystals (UD Bi-2223) annealed under conditions similar to UD2 Bi-2223 under various magnetic fields ($B \parallel c$) (Fig. \ref{fig2}). As shown in the figure, $\chi$ did not exhibit hysteresis (i.e., it is reversible) for a wide temperature range (an enlarged view of the data under magnetic fields above 100 Oe is presented in Fig. \ref{figS7} in the Appendix F). Thus, similar to Bi-2212, UD Bi-2223 had a pancake vortex that reflected the 2D nature of the electronic system, and vortex pinning was extremely weakened \cite{Blatter94}. Because the temperature at which the magnetic susceptibility was first observed to exhibit no hysteresis is the irreversible temperature ($T_{irr}$), the measured magnetic field at that temperature is referred to as the irreversible magnetic field ($B_{irr}$).

The temperature dependence of $B_{irr}$ is plotted in Fig. \ref{fig3} on the vertical axis at the log-scale. For comparison, data for optimally doped (OPT) Bi-2212 \cite{Schilling93} are also plotted. Plots with other typical copper oxide high-$T_c$ superconductors are presented in Fig. \ref{figS8} in the Appendix G. Both data exhibited an approximately straight line at high magnetic fields; However, they deviated from a straight line at low magnetic fields. This may be attributed to the crossover of the pancake vortices from a 2D state with decoupling between planes at high fields to a 3D flux lattice state at low fields \cite{Blatter94}. Moreover, the deviation from a straight line differed in case of OPT Bi-2212 and UD Bi-2223. The former bent upward whereas the latter bent downward. This is believed to be owing to differences in the anisotropic parameter $\gamma = (m_c/m_{ab})^{1/2}$, where $m_c$ and $m_{ab}$ are the effective quasiparticle masses for tunneling between and motion in the planes. The magnetic field ($B_{cr}$) that crosses two to three dimensions is theoretically expressed as $B_{cr} \approx \Phi_0/(s\gamma)^2$, where $\Phi_0$ is the flux quantum and $s$ is the distance between the conduction planes \cite{Vinokur90}. Further, $B_{cr}$ is defined as the magnetic field at which $B_{irr}$ starts deviating from the high-field linear behavior. From Fig. \ref{fig3}, $B_{cr} (\gamma)$ is $B_{cr}^{2212}$ $\approx$ 1000 Oe ( $\approx$ 100) and $B_{cr}^{2223}$ $\approx$ 20 Oe ( $\approx$ 550) for OPT Bi-2212 and UD Bi-2223, respectively. Although the anisotropy of UD Bi-2223 was consistent with the tendency of reported values near the optimal doping for Bi-2223 \cite{Piriou08} [ that is, $B_{cr} (\gamma)$ is $\approx$ 1000 Oe ($\approx$ 80), $\approx$ 600 Oe ($\approx$ 100), and $\approx$ 300 Oe ($\approx$ 140) for overdoped, optimally doped, and slightly underdoped samples, respectively.], the value appears considerably large.

The anisotropy of cuprate superconductors is generally dependent on doping. The primary reason for this is believed to be the inhomogeneity of the superconducting order parameter in \textbf{k}-space \cite{Suzuki12}. In the underdoped state, the large pseudogap opens along the anti-nodal direction \cite{Hüfner08}, resulting in a partial superconducting gap \cite{Norman98, Suzuki12}. Consequently, interlayer coupling is suppressed, thereby yielding a large $\gamma$. Moreover, the significantly large $\gamma$ observed in UD Bi-2223 may have been caused by the"nodal metallic state" of the inner CuO$_2$ plane (IP), which was recently demonstrated using angle-resolved photoemission spectroscopy (ARPES) \cite{Ideta}. Quasiparticles along the nodal direction with robust superconductivity have zero probability of inter-plane tunneling  \cite{Ioffe98}. Therefore, the nodal metallic state of IP may have decoupled the interaction between IP and the outer CuO$_2$ planes (OP), which resulted in the significantly large $\gamma$.

As mentioned above, in the 2D vortex state on the low-temperature, high-field side, $B_{irr}$ exhibits the following relationship:
\begin{equation}
%\[
B_{irr} (T) = B_{0}e^{-T/T_{0}}, 
%\]
\end{equation}
over a wide temperature range in a material-independent manner, where $B_{0}$ and $T_{0}$ are constants. However, on the lowest-temperature side, $B_{irr}$ rapidly increases toward the upper critical magnetic field ($B_{c2} (0)$). The parameter values were obtained by fitting Eq. (1) to the appropriate temperature range for each material. The results are listed in Table I in the Appendix H. The fitting was good for all the materials. Therefore, a universal property of copper oxides is that $B_{irr}$ follows Eq. (1) at low temperatures and high magnetic fields.

\section{IV. DISCUSSION}
Here, we discuss the origin of the KT transition-like phenomena. Bulk cuprate high-$T_c$ superconductors can be considered as a system of stacked single films with Josephson coupling. Thus, the system can be described by the 3D-XY model \cite{Hikami80}.  In this case, vortex/anti-vortex excitations occur at each conduction plane below $T_{c0}$, and the superconducting 3D order at $T_c$ occurs above $T_{KT}$. Thus, if $T_{KT}$ $\approx$ $T_c$ $\ll$ $T_{c0}$, it can be considered as KT transition-like \cite{Matsuda93}. According to this theory, the relationship between $T_{KT}$ and $T_c$ was formulated by \cite{Hikami80}, that is $T_c = T_{KT} + T_{KT}({\pi}/{\ln\gamma})^{2}$.
%\begin{equation}
%\[
%T_c = T_{KT} + T_{KT}\bigg(\frac{\pi}{\ln\gamma}\bigg)^{2}, 
%\]
%\end{equation}
We tentatively substituted the parameters obtained for UD1 Bi-2223 ($T_{KT}$ = 81 K) and $\gamma$ = 550 into this equation to obtain $T_c$ = 101 K. This $T_c$ value was considerably higher than $T_{KT}$ and exceeded the observed $T_{c0}$ ( = 87 K) (Fig. \ref{fig1} (a)). Thus, the difficulty in inducing the KT transition within the 3D-XY model, as highlighted by Matsuda et al. \cite{Matsuda93}, is also true for underdoped Bi-2223 with a larger anisotropy than OPT Bi-2212.

Therefore, a model beyond the simple 3D-XY model is required. Two sources of force exist between vortices in adjacent planes: Josephson coupling and electromagnetic coupling of the current loops that form the vortices. The 3D-XY model only considers the former. In cases wherein the former is extremely small, the latter becomes dominant. A study \cite{Piriou08} found that, in Bi-2223, a crossover occurred from a Josephson--coupling--dominated overdoped state to electromagnetic--coupling--dominated underdoped state in the vicinity of optimal doping. Korshunov \cite{Korshunov90} demonstrated that the interaction of layers via an electromagnetic field stabilized the vortex lines against the formation of vortex rings and restored the KT transition. Therefore, this model better explains our observations. However, a quantitative evaluation of this model remains challenging.

Alternatively, we considered the in-plane inhomogeneity proposed by Geshkenbein \cite{Geshkenbein98} and Ikeda \cite{Ikeda06} in their analysis of $B_{irr}$. As shown below, we obtained a result that quantitatively agreed with our observations. Thus, in-plane inhomogeneity is considered as a promising microscopic origin for causing the KT transition-like phenomena in the bulk body. 

Geshkenbein et al. assumed that a high $T_c$ ( $>$ $T_{c0}$) grain of size $R$, spacing $d$, and therefore, areal density $x_G$ = $R^2/d^2$ exists in the normal conducting matrix. When high $T_c$ grains are diluted ($d$ $\gg$ $R$) and placed in a zero magnetic field, the Josephson binding energy ($E_J$) can be expressed as $E_J (T, \phi)$ = $E_J^{0}e^{-d/\xi_n}F_d (\phi)$ \cite{Geshkenbein98}; where, $\xi_n$ = $v_F/2{\pi}T$ is the coherence length of the clean limit normal metal phase and $F_d (\phi)$ is a function representing the phase dependence. The energy scale of the Josephson coupling ($E_J^0$), assuming 2D nature of the system, is thus obtained $E_J^0 \approx {R}{v_F}(p_{F}R)N_c/{2{\pi}d^2}$,
%\begin{equation}
%\[
%E_J^0 \approx \frac{R}{d}\frac{v_F}{2{\pi}d}(p_{F}R)N_c, 
%\]
%\end{equation}
where $N_c$ is the number of conduction planes over which the grains are extended. In Bi-2223, three CuO$_2$ planes form one conduction plane. In contrast, in Bi-2212, two CuO$_2$ planes form one conduction plane. $N_c$ is assumed to be $\approx$ 1. Considering that the Fermi velocity ($v_F$) is universal in cuprates \cite{Zhou03} and assuming that the $p_F$ (average of three CuO$_2$ planes) of UD Bi-2223 is not significantly different from that of OPT Bi-2212, the dominant factor of $E_J^0$ is thus obtained as $x_G$ (= $R^2/d^2$). Here, we define $T_0$ = $v_F⁄2{\pi}d$, then, $E_J (T, \phi)$ = $E_J^{0}e^{-T/T_0}F_d (\phi)$. $T_c$ is expressed as,
\begin{equation}
%\[
T_c \approx E_J (T_c ), 
%\]
\end{equation}
Because the Josephson coupling between grains is significantly suppressed in a magnetic field, a formula equivalent to Eq. (1) was derived for $T$ $>$ $T_0$ \cite{Geshkenbein98}.

Figure \ref{fig3} indicates that the slopes (and thus $T_0$; further details are presented in Table I of the Appendix H) of $B_{irr} (T)$ for OPT Bi-2212 and UD Bi-2223 were nearly identical. This facilitates to the estimation of $T_c$ for UD Bi-2223 within the framework of the model described above, considering OPT Bi-2212's $T_c$  (= 89 K).  Assuming that superconducting grains are good conductors and normally conducting grains are highly resistive, as well as the distribution of inhomogeneity between these compounds being similar owing to these crystal structures having a common block layer, we can approximate $x_G$ as $x_G$ $\propto$ $\sigma_c$ \cite{Yamada03}. From the experimental results, $\rho_c$ of UD Bi-2223 was 85 ${\Omega}$cm, immediately above $T_c$ (sample a in Fig. 3 of Ref. \cite{Fujii02}), and that of OPT Bi-2212 is 7 ${\Omega}$cm (sample $\delta$ = 0.25 in Fig. 2 of ref. \cite{Watanabe00}). thus we obtain $x_G$ (thus, $E_J^0$) of UD Bi-2223 is $\approx$ 0.08 times that of OPT Bi-2212. Consequently, $T_c$ of UD Bi-2223 was determined to be $\approx$ 80 K. This value approximately agreed with the observed $T_c$ (= 82 K). Thus the shrinking of the superconducting grain and reduction in the energy of the Josephson coupling between grains, resulted in $T_c$ of UD Bi-2223 being significantly lower than its $T_{c0}$ (= 87 K). Moreover, they were comparable to $T_{KT}$ (= 81 K). Thus, it can be concluded that a KT transition-like phenomenon was observed. In OPT Bi-2212, the larger superconducting grains resulted in larger Josephson coupling energies, and superconductivity possibly occurred immediately below $T_{c0}$ \cite{Martin89}.

In fact, nanometer-sized inhomogeneities in the electronic state within the CuO$_2$ plane have been reported in recent years using scanning tunneling microscopy \cite{Pan01, Lang02, Gomes07, Kasai09, Hamidian16, Du20}. The gap size was found to be spatially distributed, with regions exhibiting small superconducting gaps. Whereas, regions showing large gaps were in a pseudogap state and were normally conducting. The region exhibiting the superconducting gap was found to shrink as the sample became more underdoped \cite{Lang02}, consistent with the numerical analysis using $\sigma_c$ described above. However, the direct in-plane Josephson properties have rarely been reported \cite{Semba00}.

In the previous reports, $T_{KT}$ was much lower (30 and 16 K, in ref. \cite{Matsuda92} and \cite{Tranquada07}, respectively) than that in this study. What is the origin for the difference? The $T_{KT}$ is the temperature at which the energy loss owing to the dissociation of the vortex-anti-vortex pair was balanced by the gain of the entropy term of the free energy associated with the dissociation. Thus, we obtain $T_{KT}$ $\propto$ $\rho_s^{2D}$, where $\rho_s^{2D}$ is the superfluid sheet density \cite{Tinkham}. Therefore, the high $T_{KT}$ of Bi-2223 may be attributable to the larger superfluid density than that of the materials in the literature \cite{Matsuda92, Tranquada07}. The large $\rho_s^{2D}$ in Bi-2223 could be attributed to the fact that the conduction layer of Bi-2223 is composed of a set of three CuO$_2$ planes.

Here, we exploited the temperature dependence of $B_{irr}$ to estimate $B_{cr}$, and numerically analyzed $B_{irr}$ for OPT Bi-2212 and UD Bi-2223. We assumed a value close to the vortex-glass transition field $B_{g}$ of ref. \cite{Geshkenbein98}. This assumption was based on the experimental finding for YBa$_2$Cu$_3$O$_y$ \cite{Nishizaki00}. However, for accurate discussions on the equilibrium vortex-matter phase diagram in this system, measurements of thermodynamic quantities such as magnetization jumps to indicate vortex-melting \cite{Zeldov95, Kadowaki98} or of $I$-$V$ characteristics to indicate the vortex-glass transition \cite{Koch89} are required. These studies are underway.

We considered the in-plane inhomogeneity as a possible cause of the observed KT transition-like behavior. However, we do not rule out other possibilities such as the Korshunov's model \cite{Korshunov90}. Furthermore, the data used for this analysis were obtained only from UD Bi-2223. The inner CuO$_2$ plane, which is unique to Bi-2223, appeared to aid the KT transition by the virtue of extremely flat and enhancing the 2D nature \cite{Iye10, Nomura19}. However, the requirement of the inner CuO$_2$ plane remains unclear. Furthermore, the extent to which KT transition-like phenomena are universal in cuprates must be investigated by studying material and systematic doping-level dependence in the future.

\section{V. SUMMARY}
The KT transition is a topological phase transition specific to a 2D system, and is considered to not occur in bulk (3D) systems. In this study, we revealed a KT transition-like superconducting transition in underdoped Bi-2223 bulk single crystals. The mechanisms governing this phenomena were discussed in terms of extreme 2D nature and/or in-plane inhomogeneity specific to this system. Consequently, the results provided new and compelling evidence that, in copper-oxide high-$T_c$ superconductors, a phase-disordered superconductivity, wherein Cooper pairs exist but their phases do not settle, exists at temperatures higher than $T_c$. This result constrains the theoretical model on high-$T_c$ superconductivity.

\section*{Acknowledgments}
The authors acknowledge the useful discussions with R. Ikeda, Y. Matsuda, S. Adachi, and Y. Itahashi. This work was supported by JSPS KAKENHI Grant Numbers 25400349, 20K03849, and 23K03317. One of the authors (T. W.) was supported by a Hirosaki University Grant for Distinguished Researchers from fiscal years 2017 to 2018.

\begin{figure*}[t]
		\begin{center}
			
			\includegraphics[width=180mm]{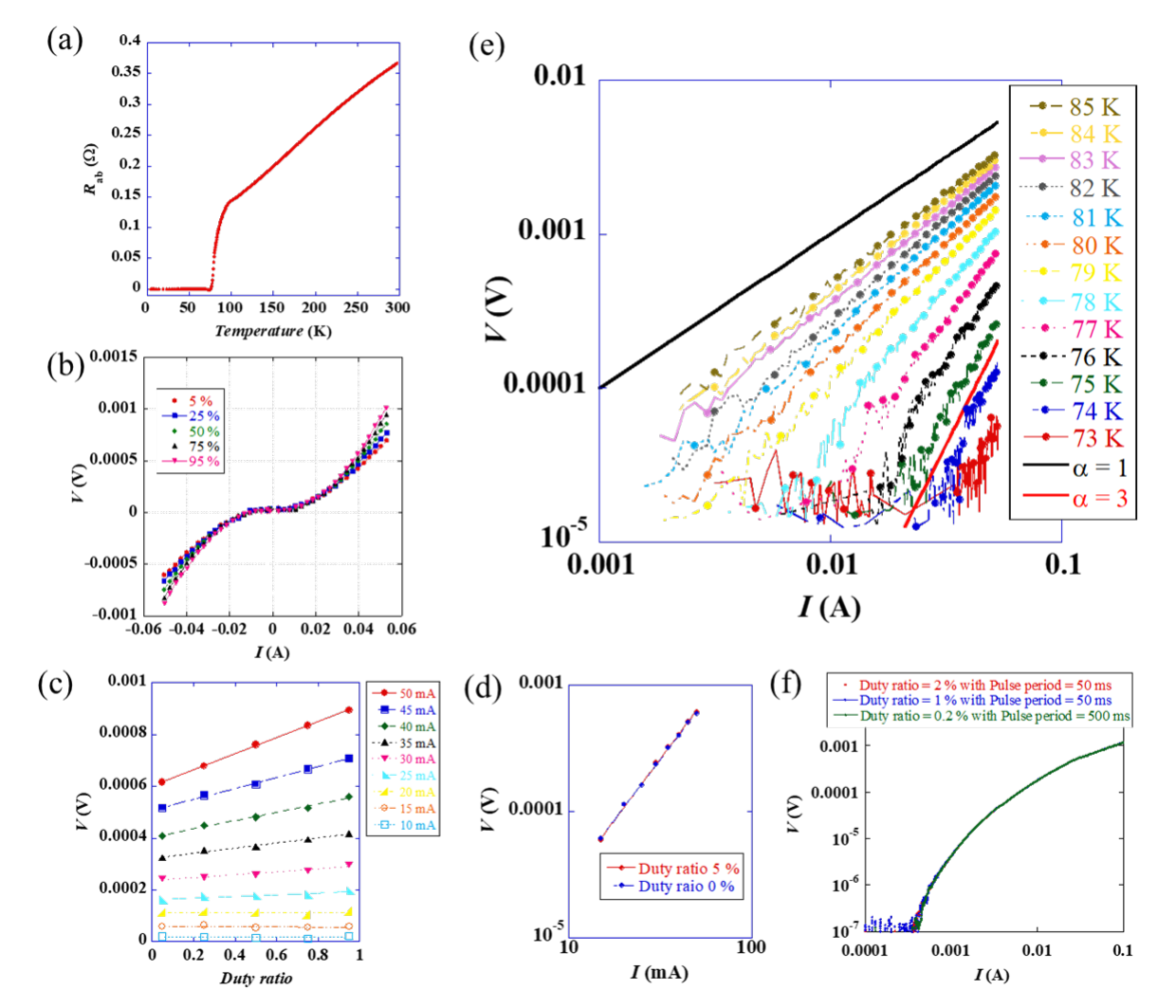}
			\caption{\label{figS1}(Color online)
				(a)Temperature dependence of the in-plane resistance ($R_{ab}$) of underdoped Bi-2223 single crystal used to examine the measurement conditions of pulsed $I-V$ characteristics ($T_c$ = 75 K). (b) Measured $I-V$ characteristics at various duty ratios at 77 K. (c) Voltage ($V$) vs. duty ratio for various current values. (d) Comparison of I-V characteristics at a hypothetical duty ratio of zero and I-V characteristics at a duty ratio of 5 \%. The two almost overlap. (e) $I-V$ characteristics near $T_c$ measured under a duty ratio of 5 \%. (f) $I-V$ characteristics measured under various conditions with extended pulse period. Measurements are performed at the superconducting state (60 K) for an underdoped Bi-2223 single crystal different from the sample shown in (a).}
		\end{center}
	\end{figure*}			

\begin{figure*}[t]
		\begin{center}
			
			\includegraphics[width=190mm]{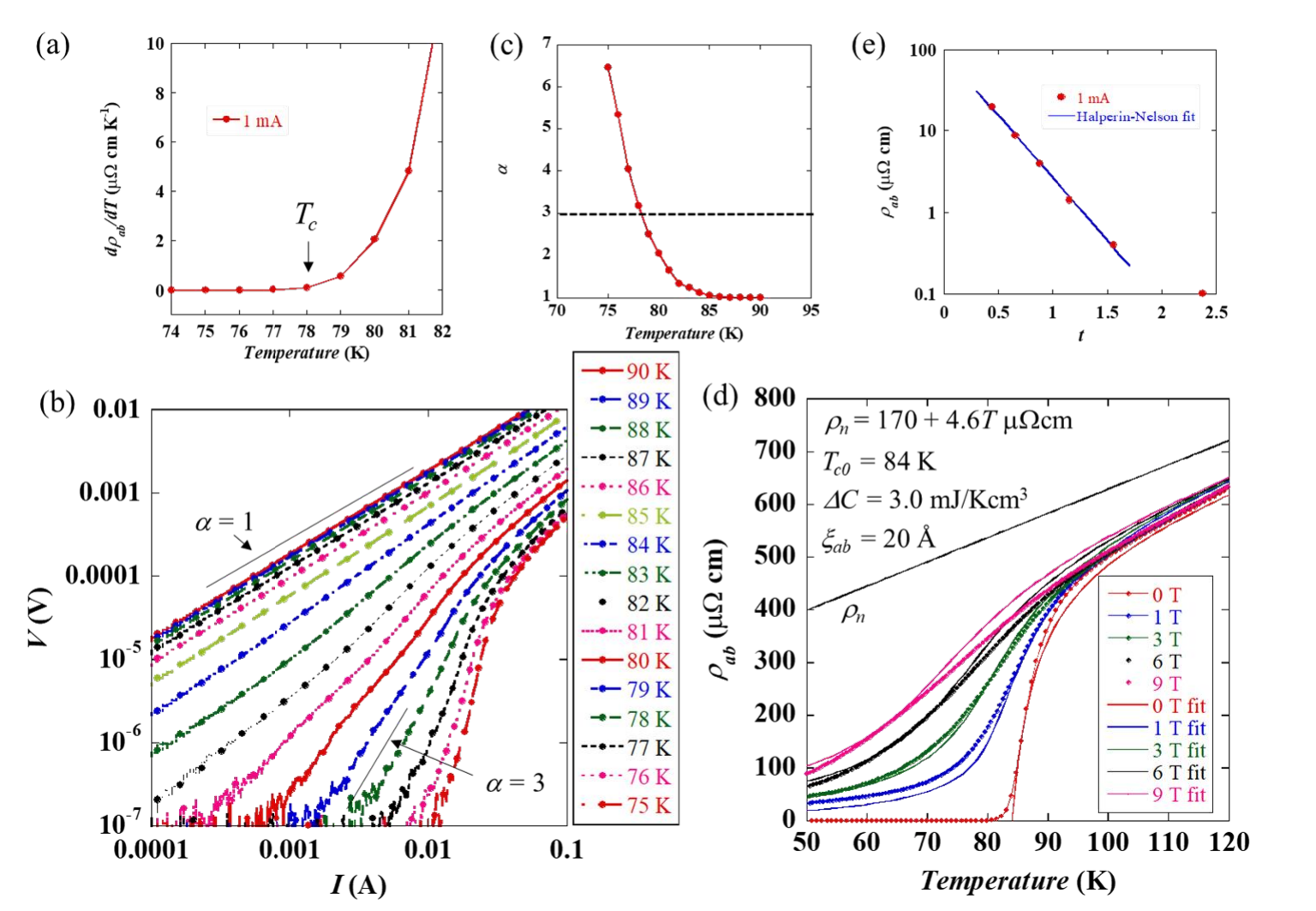}
			\caption{(Color online)
				(a) Temperature derivative of zero-field in-plane resistivity ($d\rho_{ab} (T)/dT$) vs. temperature. (b) $I$–$V$ characteristics near $T_c$ (= 78 K). (c) Temperature dependence of the power exponent $\alpha$. (d) Temperature dependence of $\rho_{ab}$ with and without magnetic fields, and their fitting results using IOT theory.  (e) Temperature dependence of $\rho_{ab}$ at $T_{KT} < T < T_{c0}$. The straight line indicates the fitting result using the Halperin-Nelson equation,  $\rho_{ab} (T)$ = $\rho_{ab}^0$$\exp(-2ct)$, where $t = [({T_{c0} - T})/(T - T_{KT})]^{1/2}$, with $\rho_{ab}^{0}$ = 90 $\mu\Omega$cm and $c$ = 1.76.}
\label{figS2}

		\end{center}
	\end{figure*}

\begin{figure}[t]
		\begin{center}
			
			\includegraphics[width=90mm]{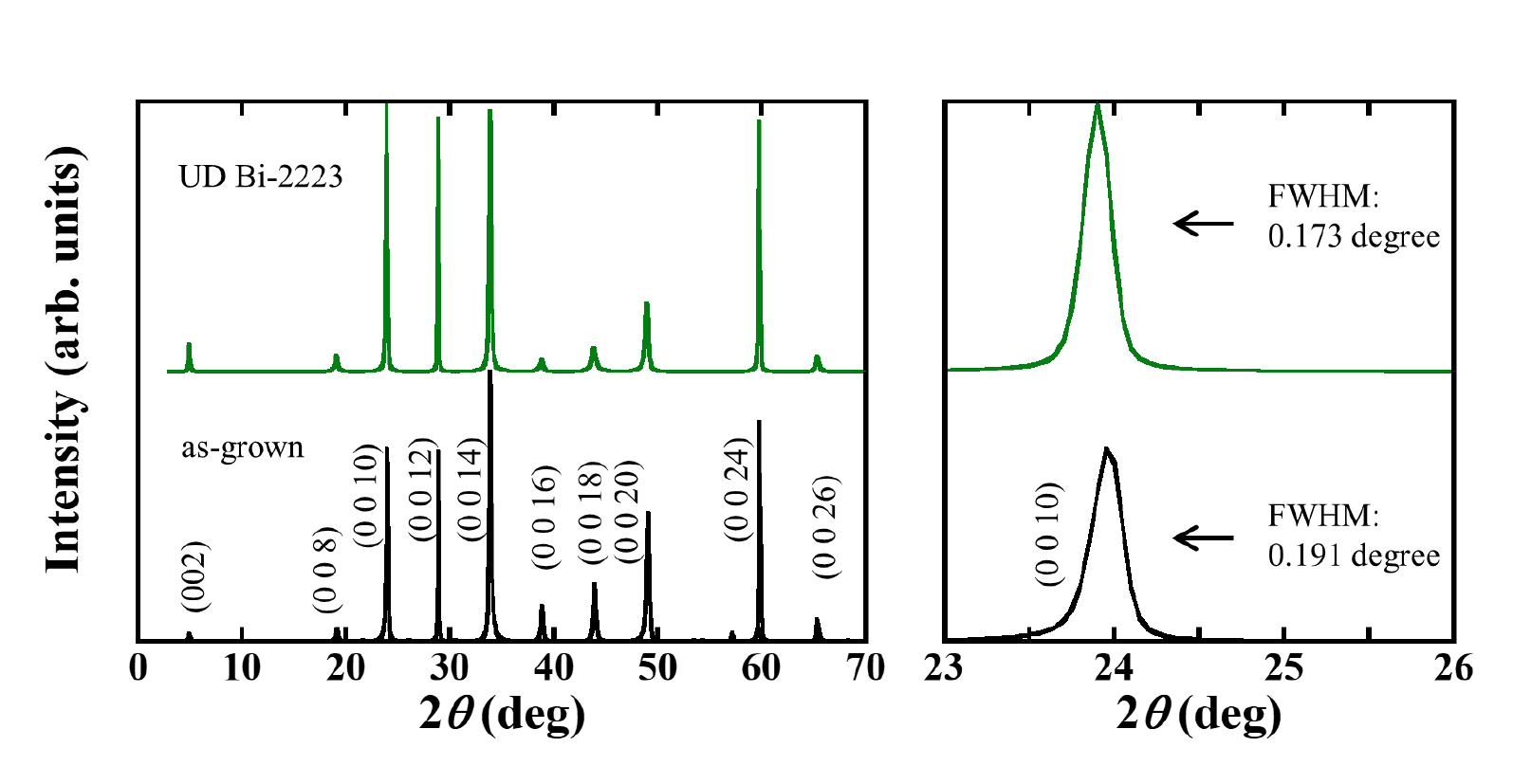}
			\caption{(Color online)
				X-ray diffraction (XRD) patterns of as-grown and UD Bi-2223 single crystals.}
\label{figS3}
			
		\end{center}
	\end{figure}

\begin{figure*}[t]
		\begin{center}
			
			\includegraphics[width=180mm]{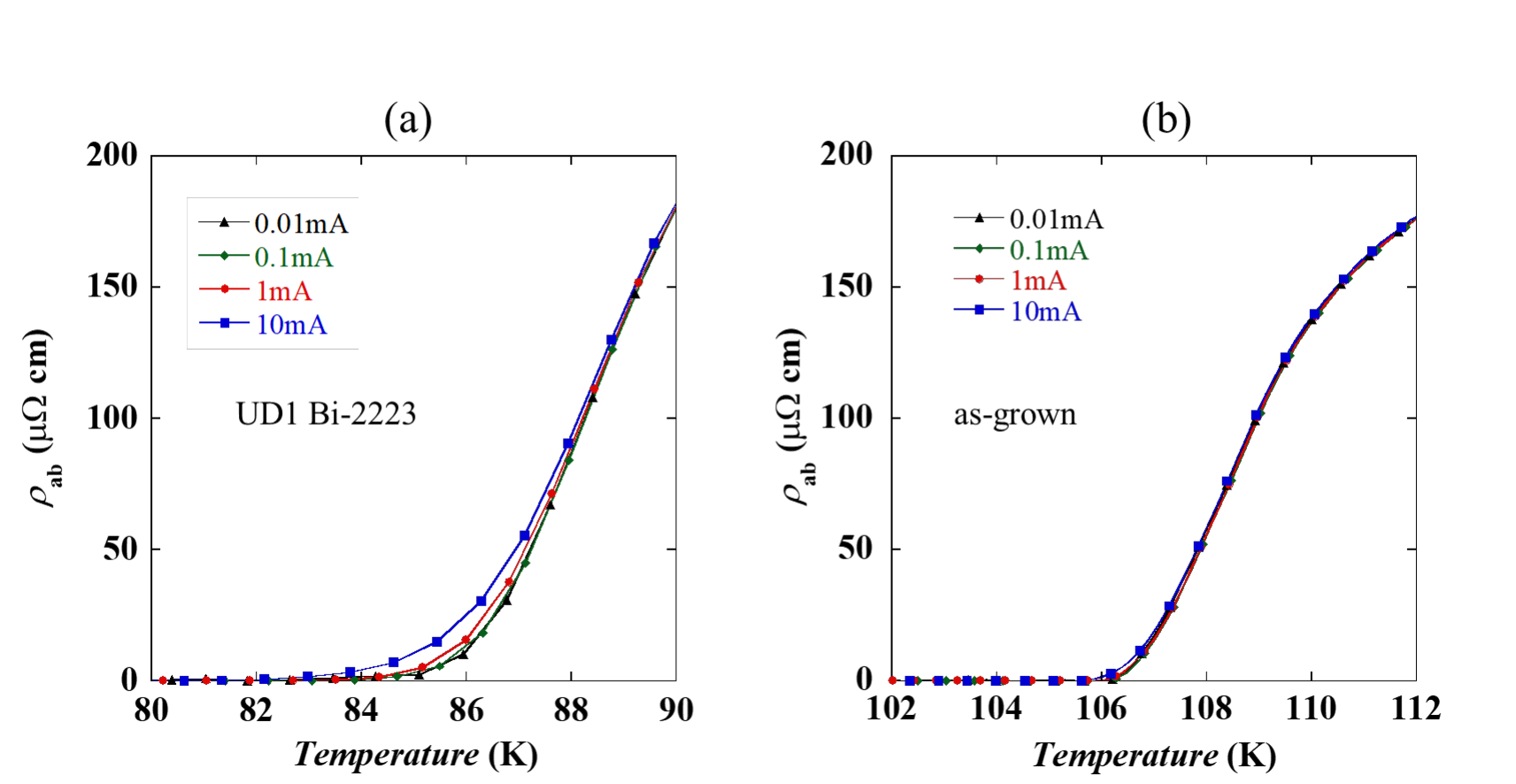}
			\caption{(Color online)
				In-plane resistivity $\rho_{ab}$ of (a) an underdoped (UD1) and (b) as-grown Bi-2223 single crystals with various applied currents.}
\label{figS4}
			
		\end{center}
	\end{figure*}

\begin{figure*}[t]
		\begin{center}
			
			\includegraphics[width=180mm]{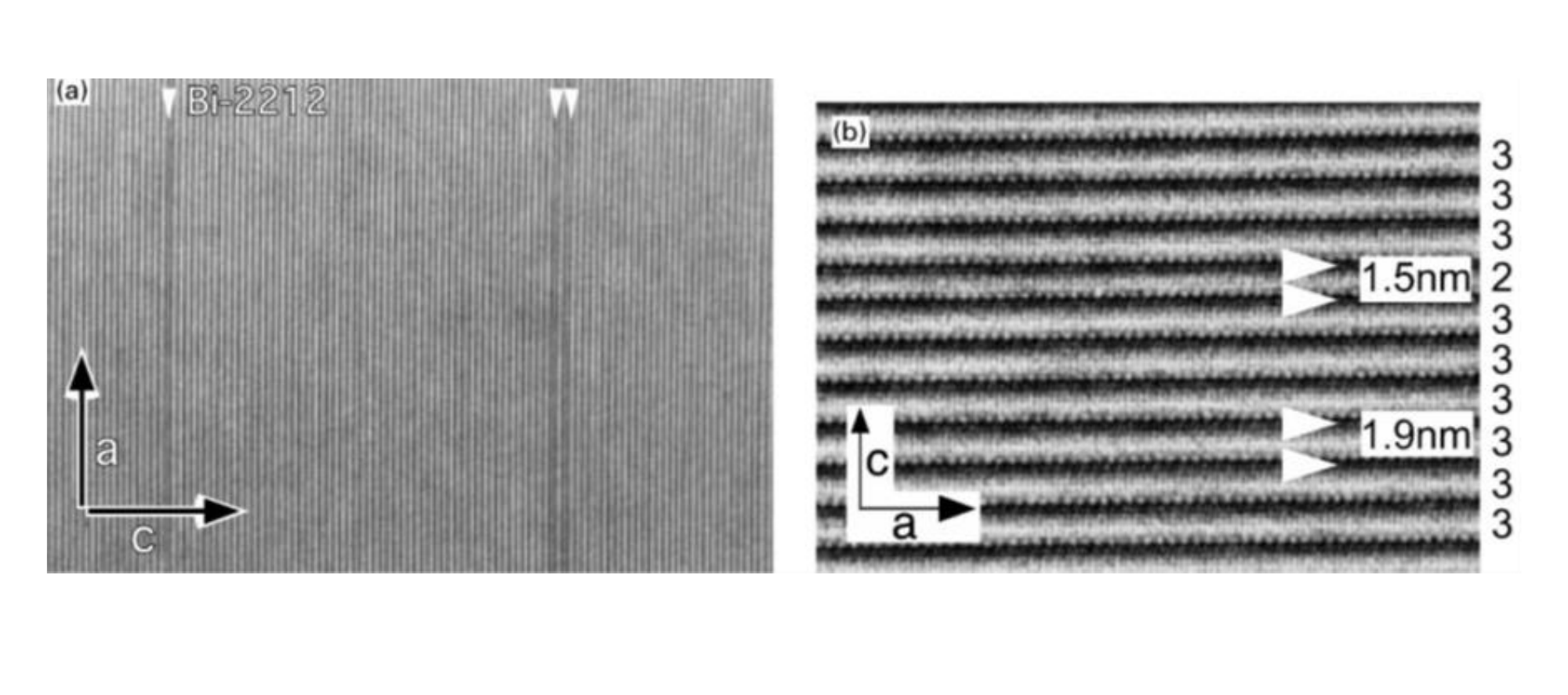}
			\caption{(Color online)
				(a) TEM image of a Bi-2223 single crystal, captured with the incident beam parallel to the $b$ direction. (b) The scale is expanded to show the intergrowth of Bi-2212. Bi-2212 marked as “2” is intergrown in the Bi-2223 marked as “3”. The data is reproduced with permission from ref. \cite{Fujii01}.}
\label{figS5}
			
		\end{center}
	\end{figure*}

\begin{figure*}[t]
		\begin{center}
			
			\includegraphics[width=180mm]{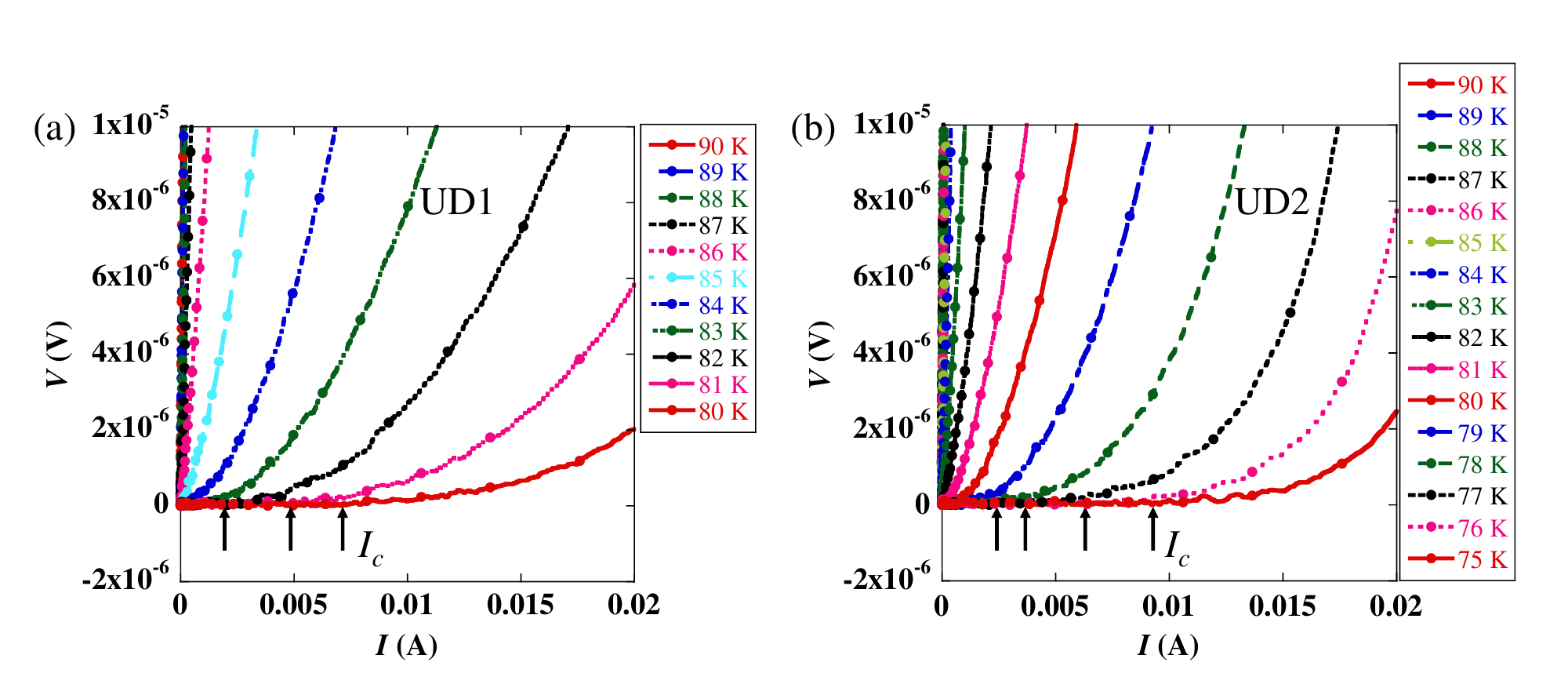}
			\caption{(Color online)
				Expanded $I$-$V$ characteristics plotted as linear scales for (a) UD1 Bi-2223 and (b) UD2 Bi-2223. The arrows indicate $I_{c}$s for each temperatures. $I_{c}$ is defined as the current at which $V$ becomes below 1 x 10$^{-8}$ V (noise level). (a) $I_{c}$ is 0.0020$\pm$0.0005, 0.0048$\pm$0.0005, and 0.0071$\pm$0.0010 A for 82 (= $T_c$), 81, and 80 K, respectively. (b) $I_{c}$ is 0.0026$\pm$0.0005, 0.0038$\pm$0.0005, 0.0063$\pm$0.0010, and 0.0092$\pm$0.0010 A for 78 (= $T_c$), 77, 76, and 75 K, respectively.}
\label{figS6}
			
		\end{center}
	\end{figure*}

\begin{figure}[t]
		\begin{center}
			\includegraphics[width=90mm]{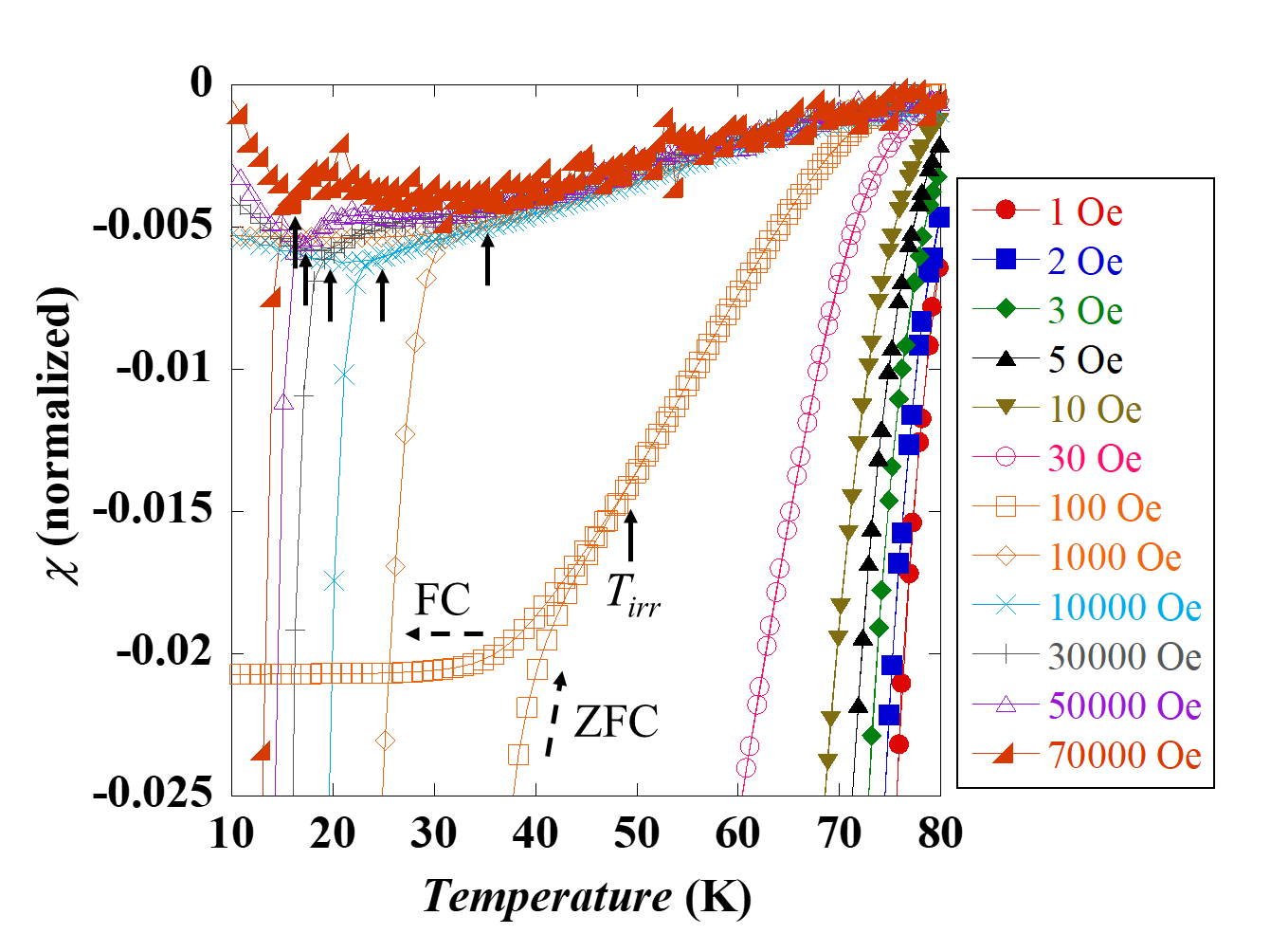}
			\caption{\label{figS7}(Color online) Enlarged view of the region of the low magnetic susceptibility in Fig. 2 in the main text. Hysteresis is observed even upon the application of a high magnetic field, and an irreversible temperature ($T_{irr}$) can be defined. The arrows indicate $T_{irr}$ under the respective applied magnetic fields.}
		\end{center}
	\end{figure}

\begin{figure}[t]
		\begin{center}
			\includegraphics[width=90mm]{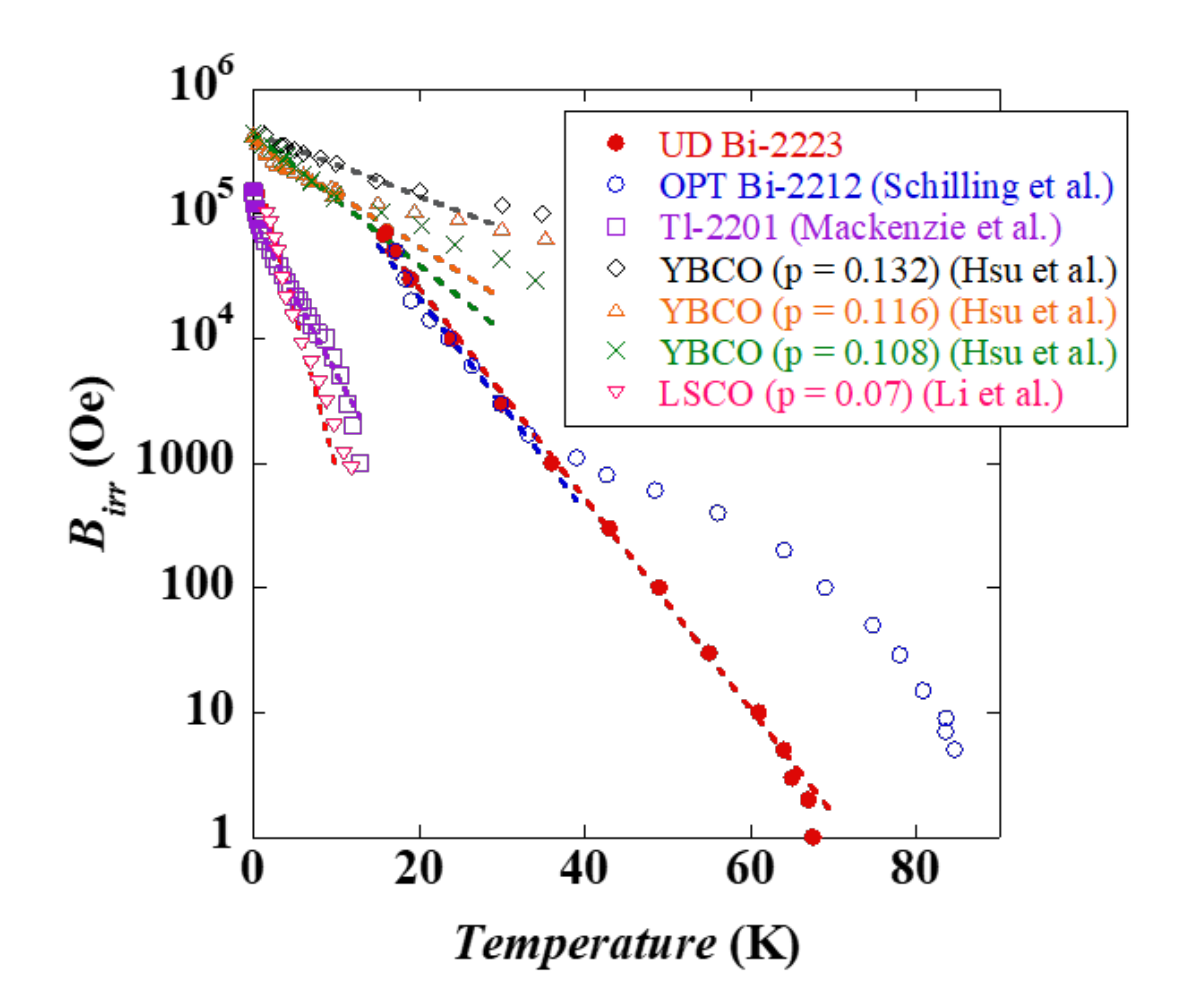}
			\caption{\label{figS8}(Color online)  Temperature dependence of the irreversible magnetic field ($B_{irr}$) for underdoped (UD) Bi-2223, optimally doped (OPT) Bi-2212 \cite{Schilling93}, Tl$_2$Ba$_2$CuO${6+\delta}$ (Tl-2201) \cite{Mackenzie93}, LSCO (p = 0.07) \cite{Li07}, and YBCO (p = 0.132, 0.116, 0.108) \cite{Hsu21}. The vertical axis is shown on a logarithmic scale. The dotted lines indicate the fitting results using the formula, $B_{irr} (T) = B_{0}e^{-T/T_{0}}$, on the low-temperature and high-field region.}
		\end{center}
	\end{figure}

%\begin{figure}[t]
%		\begin{center}
			%\includegraphics{fig01.eps}
			%\includegraphics[width=174mm]{Figure2-(usui).pdf}
%			\includegraphics[width=90mm]{Figure3.pdf}
%			\caption{\label{fig3}(Color online)Temperature dependence of the irreversible magnetic field ($B_{irr}$) for underdoped (UD) Bi-2223 and optimally doped (OPT) Bi-2212 %\cite{Schilling93}. The vertical axis is shown on a logarithmic scale. The dotted lines indicate the fitting results using the formula, $B_{irr} (T) = B_{0}e^{-T/T_{0}}$, in the low-temperature %and high-field region. The crossover fields ($B_{cr}$) from two to three dimensions are indicated via the arrows for OPT Bi-2212 and UD Bi-2223.}
%		\end{center}
%	\end{figure}
\begin{table*}

   \caption{Parameter values determined by fitting the irreversible magnetic field ($B_{irr} (T)$) for various cuprate high-$T_c$ superconductors.}
   \label{tab1}
   \vspace*{0.5cm}

\begin{tabular}{cccccccc}
\hline
\hline
  & UD Bi-2223 & OPT Bi-2212 & Tl 2201 & LSCO (p = 0.07) & YBCO (p = 0.132) &  YBCO (p = 0.116) &  YBCO (p = 0.108)  \\
\hline
$B_{0}$ (T)  & 129 & 108 & 9.2 & 28 & 44 & 38 & 44 \\
$T_{0}$ (K)  & 5.1 & 5.1 & 3.5 & 1.8 & 17 & 10 & 8.2 \\
\hline
\hline
\end{tabular} 
\end{table*}

\appendix
\section{Appendix A: Examination of pulse current($I$)-voltage($V$) characteristics measurement conditions}

Measurements of the current($I$)–voltage($V$) characteristics of a bulk sample and the determination of the power exponent $\alpha$ ($V \propto I^{\alpha}$) necessitate a current of several tens of mA. Such a large current may cause Joule heating within the sample, even in the case of pulse measurements \cite{Suzuki07}. Therefore, prior to performing this measurement, we investigated the conditions for the pulse measurement to suppress Joule heating. First, we fixed the pulse period at 1 msec and varied the duty ratio (current pulse width/pulse period) to investigate its effect.

Figure \ref{figS1}(a) shows the temperature dependence of the in-plane resistance ($R_{ab}$) of the sample. It is an underdoped sample with a superconducting transition temperature ($T_c$) of 75 K. The $I-V$ characteristics were measured at various duty ratios at 77 K, that is immediately above $T_c$ (Fig. \ref{figS1}(b)). As the duty ratio increased, the nonlinearity increased. Thus, an increase in the duty ratio caused the effect of heat generation to become more pronounced. The results were plotted against the voltage ($V$) on the vertical axis and the duty ratio on the horizontal axis for various current values (Fig. \ref{figS1}(c)). At any current value, the voltage ($V$) increased approximately linearly with the duty ratio. Therefore, the $V$ values obtained by extrapolating this straight line to a duty ratio of zero were considered as the true voltages in the absence of heat generation. The $I-V$ characteristics at a duty ratio of zero and those at a duty ratio of 5 \% are plotted in Fig. \ref{figS1}(d). The two almost overlapped, and the power exponents $\alpha$ were also approximately the same ($\alpha$ = 1.93 and $\alpha$ = 1.90  for the duty ratios of 5 \% and 0 \%, respectively). Therefore, we considered that the data with a duty ratio of 5 \% sufficiently suppressed the heat generation. Figure \ref{figS1}(e) shows the $I-V$ characteristics near $T_c$, measured under this condition. On the high-temperature side, it was approximately Ohmic ($\alpha$ = 1); however, $\alpha$ increased as the temperature decreased, reaching $\alpha$ = 3 near $T_c$ (= 75 K).

However, as shown in Fig. \ref{figS1}(e), the data on the low current side were slightly noisy. Therefore, to improve the data signal to noise (S/N) ratio, we extended the pulse period while maintaining the duty ratio below 5 \%.  Specifically, the $I-V$ characteristics were examined by varying the pulse period across 50-500 ms and the duty ratio across 0.2-2 \% (Fig. \ref{figS1}(f)). Because there was no difference in the data under any condition, the effect of heat generation was considered negligible under these conditions. Consequently, the measurement conditions were set to a pulse period of 50 ms and duty ratio of 2 \%.

\section{Appendix B: Transport properties of the UD2 Bi-2223 sample}

Figure \ref{figS2}(a) shows the temperature dependence of the temperature derivative of the zero-field in-plane resistivity $d\rho_{ab} (T)/dT$ for the UD2 Bi-2223 sample.  Consequently, we determined $T_c$ = 78 K, according to the definition in the text.

Figure \ref{figS2}(b) shows its $I-V$ characteristics. On the high-temperature side, it was approximately Ohmic ($\alpha$ = 1); however, $\alpha$ increased with decreasing temperature, reaching $\alpha$ = 3 near $T_c$ (= 78 K) ($T_{KT}$  = 78 K). Figure \ref{figS2}(c) shows the temperature dependence of $\alpha$ thus obtained. Notably, the $I$–$V$ characteristics were $S$-shaped below 84 K. The S-shaped behavior has already been presented in the main text (Fig. 1(c)); however, the behavior was more pronounced (Fig. \ref{figS2}(b)). This behavior is similar to that of 2D MoS$_2$ with carrier injection owing to the electric field effect \cite{Saito20}. In the low-current regime, the voltage increased significantly with an increase in the dissociated vortex and antivortex pairs as the current increased. As the current was further increased, a phase slip line formed, the viscous force against the vortex (anti-vortex) weakened, the velocity of the vortex (anti-vortex) increased, and the voltage increased further \cite{Weber91}. However, at higher currents, the vortex (anti-vortex) disappeared and approached the $I$–$V$ characteristics of the normal state. Therefore, the $S$-shaped behavior was considered a characteristic of systems that undergo KT transition. The observation of a behavior similar to that of two-dimensional MoS$_2$ with a certain KT transition indicated that the UD1 and UD2 Bi-2223 used in this study also cause the excitation of vortices and anti-vortices similar to the KT transition.

Figure \ref{figS2}(d) shows the temperature dependence of the in-plane resistivity ($\rho_{ab}$) with and without magnetic fields and the fitting results based on theory. An analysis of the zero-magnetic-field data revealed the mean-field superconducting transition temperature ($T_{c0}$) as 84 K. The data in the magnetic field were analyzed using the theory formulated by Ikeda-Ohmi-Tsuneto ( IOT theory) \cite{Ikeda91}. We previously performed a similar analysis for Bi-2223 and Bi-2212 near the optimum doping (further details on the analysis method can be found in \cite{Adachi151}). Because the theoretical formula includes the specific heat jump ($\Delta$$C$), in-plane and inter-plane coherence lengths ($\xi_{ab}$ and $\xi_{c}$, respectively), and $T_{c0}$ as parameters, they can be evaluated by fitting the transition curves in the magnetic fields. Here, fitting was performed with $\xi_{c}$ fixed at $\xi_{c}$ = 0.1 Å because even if $\xi_{c}$ was reduced further, the fitting curve did not change. From the fitting, $\xi_{ab}$ was obtained as $\approx$ 20 Å. Thus, the anisotropy parameter ($\gamma$) was determined as $\gamma$ = $\xi_{ab}/\xi_{c}$ $\ge$ 200. This result was consistent with the value of $\gamma$ ($\approx$ 550) obtained by evaluating the crossover magnetic field ($B_{cr}$) in the main text. Furthermore, $\Delta$$C$ $\approx$ 3.0 mJ/Kcm$^3$ was obtained. This value was considerably smaller than that of optimally doped Bi-2212 ($\Delta$$C$ $\approx$ 20.0 mJ/Kcm$^3$) \cite{Adachi151}. The superfluid density ($\rho_s$) (phase stiffness) is related to the magnetic field penetration depth ($\lambda_L$) as $\rho_s$ $\propto$ $1⁄{\lambda_L}^2$. Further, $\lambda_L$ is related to $\Delta$$C$ as $1⁄{\lambda_L}^2$ = $32\pi^3\xi_{ab}^2T_{c0}{\Delta}C⁄{\Phi_0}^2$ (where $\Phi_0$ is the magnetic flux quantum) \cite{Ikeda91}. Using these relations, we estimated $1⁄{\lambda_L}^2$ as 23.4 ${\mu}m^{-2}$ and 44.1 ${\mu}m^{-2}$ for UD2 Bi-2223 and optimally doped Bi-2212 \cite{Adachi151}, respectively. Thus, underdoped Bi-2223 has a smaller $\rho_s$ than optimally doped Bi-2212. This renders it difficult for Cooper pairs to be coherent in phase, and thus underdoped Bi-2223 is more likely to undergo KT transition-like phenomena.

Figure \ref{figS2}(e) presents the tailored temperature dependence of $\rho_{ab}$ below $T_{c0}$ (84 K). The data are approximately on a straight line. Thus, as shown in Fig. 1(e) in the main text, the tailed $\rho_{ab}$ was owing to the free-vortex and anti-vortex excitations.

\section{Appendix C: Information on the quality of the samples}

Figure 6 shows the X-ray diffraction patterns of an as-grown sample and an underdoped sample (UD Bi-2223) prepared under the same annealing conditions as that in the case of UD2 Bi-2223. They are obtained from the same Bi-2223 crystal rod used in this study. In both cases, only the peak of (0 0 2n) of Bi-2223 was observed; thus, it had a high-purity Bi-2223 single crystal with little contamination of Bi-2212. As the crystal is a sample with excellent orientation, there is no contamination with other impurities. Owing to the full width at half maximum (FWHM) being narrower in the annealed sample than in the as-grown sample, it can be concluded that there is no problem of macroscopic oxygen inhomogeneity associated with annealing. The narrowing of the FWHM with annealing is reproducible \cite{Adachi15}.

Figure 7 shows the temperature dependence of the resistivity $\rho_{ab}$ of (a) UD1 Bi-2223 and its (b) as-grown crystal with various bias currents. For comparison, the temperature on the horizontal axis was plotted at the same width, and the resistivity on the vertical axis was plotted on the same scale. In the UD1 Bi-2223 crystal, a noticeable tailing behavior was evident. Further, the tailing behavior was enhanced in magnitude at the temperature intervals with increasing the bias currents. The non-ohmic resistance cannot be explained by the non-uniformity of the crystal. Furthermore, in the as-grown crystal, the tailing behavior and the non-ohmicity were suppressed. The FWHM of the X-ray diffraction pattern implied that the crystallinity of the as-grown crystal was more disordered than that of the annealed crystal. The suppressed tailing behavior and the non-ohmicity in the disordered as-grown crystal strongly indicated that the anomalies observed in the UD1 Bi-2223 crystal were not due to crystalline inhomogeneity. 

Figure 8 shows a transmission electron microscope (TEM) image of a Bi-2223 single crystal wherein only the (0 0 2n) peaks of Bi-2223 were observed in the XRD pattern, as in Fig. 6 (and therefore had the same quality as the sample used in this study), taken with an electron beam incident in the $b$-axis direction \cite{Fujii01}. No stacking faults were observed. A slight (approximately 2 \%) intergrowth of Bi-2212 was observed. However, these intergrowths did not affect the results of this study. This is because the $T_c$ of the Bi-2223 is considerably higher than that of the Bi-2212.

\section{Appendix D: Why we attribute the change of  $\alpha$ from 1 to 3 to the KT transition}

We have observed the change of $\alpha$ from one to three in Fig. 1(d) in the main text. We conclude that this phenomenon is owing to the KT transition as follows.

Similar $I$-$V$ characteristics have been reported in cuprate superconductors \cite{Koch89} However, this phenomenon pertains specifically to the vortex glass transition in a magnetic field, and fundamentally differs from the zero-field phenomenon being investigated in this study. Further, it has been reported that the $I$-$V$ characteristics follow a power-law behavior at the glass transition temperature ($T = T_g$); however, the power exponent $\alpha$ is not necessarily three (for YBCO, 2.5 \textless $\alpha$ \textless 3.0 \cite{Blatter94}, and for Bi-2223, $\alpha$ \textless 1.0 \cite{Yu23}. Furthermore, this power-law behavior does not hold in the temperature range above and below the transition temperature.

For superconducting practical wires, the power exponent $n$-value of the $I$-$V$ characteristic ($V \propto I^{n}$) indicates how rapidly the $I$-$V$ curve increases after exceeding the critical current value ($I_c$ $\approx$ 50-100 A) at temperatures below $T_c$. However, under normal conditions (in the absence of serious damage owing to bending stress, tensile strain, etc.), $n$ $\approx$ 10 -- 60 \cite{Ochiai12}.

Therefore, in this study, the change in $\alpha$ from one to three near $T_c$ under zero magnetic field is sufficient evidence for the occurrence of KT transition without any other interpretation.

\section{Appendix E: Appearance of a critical current ($I_c$) and evolution of $\alpha$ below $T_c$}

For convenience, we estimated $\alpha$ via power law fitting for the $I$–$V$ characteristics ($V \propto I^{\alpha}$) below $T_c$ in the same manner as above $T_c$ (see, Fig. 1(d) in the main text and Fig. 5(c) in the Appendix B). However, notably, the obtained $\alpha$ below $T_c$ was only an approximate value for the following reasons:

Figure \ref{figS6}(a) and \ref{figS6}(b) show the $I$–$V$ characteristics in the low bias current ($I$) and low voltage ($V$) regimes for the UD1 Bi-2223 and UD2 Bi-2223 samples, respectively. As evident, $I_c$ appeared below $T_c$. Therefore, the power-law fitting was not exact. Thus, the fact that $\alpha$ reached three below or near $T_c$ does not necessarily imply the occurrence of a true KT transition. This may have resulted in that $\alpha$ does not "jump" to 3 in this study. Here, the 3D superconducting transition occurred at $T_c$,  immediately above  the "hypothetical" $T_{KT}$ at which $\alpha$ reached 3.

\section{Appendix F: Magnified view of temperature dependence of magnetic susceptibility}

Figure \ref{figS7} shows an enlarged view of the region with low magnetic susceptibility to facilitate the easier observation of the high-magnetic-field data in Fig. 2 in the main text. As evident, hysteresis was observed even upon the application of a high magnetic field, and the irreversible temperature ($T_{irr}$) could be defined. However, the data for cooling in a magnetic field (FC) changed in the opposite direction (increasing with decreasing temperature). The cause of this phenomenon is not well-understood.

\section{Appendix G: $B_{irr}$ vs. $T$ with other typical copper oxide high-$T_c$ superconductors}

Figure \ref{figS8} presents plots of $B_{irr}$ vs. $T$ for typical copper oxide high-$T_c$ superconductors, UD Bi-2223 (this study), optimally doped (OPT) Bi-2212 \cite{Schilling93},Tl$_2$Ba$_2$CuO${6+\delta}$ (Tl-2201) \cite{Mackenzie93}, LSCO (p = 0.07) \cite{Li07}, and YBCO (p = 0.132, 0.116, 0.108) \cite{Hsu21}.  Here, irreversible fields and vortex-lattice melting fields were not distinguished. Moreover, the methods for determining $B_{irr}$ varied, depending on the literature. Nevertheless, all data were almost on a straight line at high magnetic fields, but deviated from a straight line at low magnetic fields. Therefore, this activation type behavior for $B_{irr}$ may be universal in cuprates.

\section{Appendix H: Parameter values determined by fitting the irreversible magnetic field ($B_{irr} (T)$)}%

Table I lists the parameter values determined by fitting the irreversible magnetic field ($B_{irr} (T)$) of various cuprate high-$T_c$ superconductors exhibited in Fig. \ref{figS8}. The fitting was performed using the formula  $B_{irr} (T) = B_{0}e^{-T/T_{0}}$.  UD Bi-2223 and OPT Bi-2212 had the same $T_{0}$.

%\begin{acknowledgments}

%\end{acknowledgments}

%\newpage %Just because of unusual number of tables stacked at end
%\bibliography{Referencefile1}% Produces the bibliography via BibTeX.
%\bibliography{Referencefile2}

\end{document}